\documentclass[aps,prl,twocolumn,superscriptaddress,raggedbottom,showpacs,floatfix]{revtex4}
\usepackage{amsmath,amsfonts,amssymb,amsthm}
\usepackage{bm}
\usepackage{color}
\usepackage{graphicx,latexsym}
\usepackage{epstopdf}
\usepackage{comment}
\usepackage{bm}% bold math

\def\SMB{SmB$_{6}$ }
\def\RuO{RuO$_{2}$ }

\begin{document} 
\renewcommand{\vec}{\mathbf}
\renewcommand{\Re}{\mathop{\mathrm{Re}}\nolimits}
\renewcommand{\Im}{\mathop{\mathrm{Im}}\nolimits}

\title{Direct Observation of Surface State Thermal Oscillation in \SMB Oscillators}
\author{Brian Casas}
\affiliation{Department of Physics and Astronomy, University of California, Irvine, Irvine, California 92697, USA.}
\author{Alex Stern}
\affiliation{Department of Physics and Astronomy, University of California, Irvine, Irvine, California 92697, USA.}
\author{Dmitry K. Efimkin}
\affiliation{Department of Physics, University of Texas, Austin, Austin, Texas, 78712, USA}
\author{Zachary Fisk}
\affiliation{Department of Physics and Astronomy, University of California, Irvine, Irvine, California 92697, USA.}
\author{Jing Xia}
\affiliation{Department of Physics and Astronomy, University of California, Irvine, Irvine, California 92697, USA.}

\date{\today}

\begin{abstract}

\SMB is a mixed valence Kondo insulator that exhibits a sharp increase in resistance following an activated behavior that levels off and saturates below 4K. This behavior can be explained by the proposal of \SMB representing a new state of matter, a Topological Kondo insulator, in which a Kondo gap is developed and topologically protected surface conduction dominates low temeprarture transport. Exploiting its non-linear dynamics, a tunable \SMB oscillator device was recently demonstrated, where a small DC current generates large oscillating voltages at frequencies from a few $\hbox{Hz}$ to hundreds of $\hbox{MHz}$. This behavior was explained by a theoretical model describing the thermal and electronic dynamics of coupled surface and bulk states. However, a crucial aspect of this model, the predicted temperature oscillation in the surface state, hasn't been experimentally observed to date. This is largely due to the technical difficulty of detecting an oscillating temperature of the very thin surface state. Here we report direct measurements of the time-dependent surface state temperature in \SMB with a \RuO micro-thermometer. Our results agree quantitatively with the theoretically simulated temperature waveform, and hence support the validity of the oscillator model, which will provide accurate theoretical guidance for developing future \SMB oscillators at higher frequencies. 

\end{abstract}

\maketitle

Samarium Hexaboride (SmB$_6$), a strongly correlated mixed valence compound\cite{Fisk96, Menth69}, has garnered much attention recently due to the discovery of a low temperature conducting surface state \cite{Wolgast12, Zhang13, Kim12, Li13,Kim11,Kim13}. It is proposed to form a strongly interacting version of a Topological Insulator \cite{Dzero10, Dzero12, Alexandrov13,Dzero16,Lu13}, with a Kondo Gap in the bulk and a gapless states on the surface that seems to be protected by time-reversal symmetry \cite{Kim13}. The lack of impurity conduction in the bulk has allowed the realization of a radio frequency oscillator device \cite{Stern16} based on \SMB micro-crystals, where a small DC current drives the \SMB crystal into an oscillating state and generates large oscillating voltages. 

This behavior was explained by a theoretical model \cite{Stern16}, which utilized a coupling between two conduction channels, namely, metallic surface conduction and thermally activated bulk conduction which result in significant differences in sample heating under an applied current. While this Letter will focus on the verification of this model in a proposed topological Kondo insulator, a general seiconducting sample with metallic channels with electrical and thermal coupling can be described equally well. The proposed model can be written as follows:

\begin{equation}
\begin{split}
CR_\mathrm{S} \dot{I}_\mathrm{S}=I_0-G I_\mathrm{S}; \\ C_\mathrm{H} \dot{T} =2I_\mathrm{S}^2 R_\mathrm{S} G-I_\mathrm{S} I_0 R_\mathrm{S}-\gamma(T-T_B ).
\label{Model}
\end{split}
\end{equation}

Here $I_S$ and $I_0$ are the surface and applied currents through the sample, respectively; C is the total capacitance representing the combined internal and external capacitance ; G = ($R_\mathrm{S}$ + $R_\mathrm{B}$)/$R_\mathrm{B}$ where $R_\mathrm{S}$ is the surface resistance and $R_\mathrm{B}=R_\mathrm{B}^0 \mathrm{exp}[-\Delta/T+\Delta/T_0]$ is the bulk resisatnce. $R_\mathrm{B}^0$ is bulk resistance in equilibrium at temperature $T_0$. $C_\mathrm{H}^0$ is the heat capacity and $\gamma$ is a temperature independent heat transfer rate through external leads.

Solving this simultaneous set of equations yields oscillatory solutions of both temperature and crystal voltage which has a constant phase shift of 90\textsuperscript{o}. Qualitatively one can understand the production of thermal and voltage oscillations as arising from alternating the dominating conduction channels between surface and bulk dominated conduction. Due to the Joule heating that occurs while surface conduction dominates the crystal may heat locally enough to enter a lower resistance bulk dominated state. It is in this state that Joule heating is reduced, the sample temperature will equilibrate back towards that of the measurement stage and thus drive the sample back to surface dominated conduction. This model gives predictions of oscillating voltages that agree with experiments. 

The verifaction of this model is crucial for the search and developement of novel materials and structures that can be used to improve existing oscillator technology and to realize reliable, stable, high frequency oscillation devices. Since the design of higher frequency oscillator up to THz \cite{Stern16} relies on the validity of above model, it is vitally important to verify experimentally the other key component of the model, the oscillating surface temperature. The difficulty in observing these thermal oscillations is multifaceted. Primarily, the magnitude of the temperature oscillations is predicted to be small, on the order of a Kelvin. Secondly the thermometer needs to be small enough to measure the surface state. Additionally, due to the frequencies of the thermal oscillations the measurement method needs to be able to equilibrate to the surface temperature quickly, as to not introduce a significant contribution to a phase shift between voltage and thermal oscillations. Here we have overcome these challenges with a \RuO micro-thermometer and fast data acquisition methods.

The studied devices were constructed of a single crystal of SmB$_6$, held at 2K in a cryostat under no applied field, in parallel with an external capacitor stored at room temperature outside of the cryostat. A detailed schematic of this device can be seen in Fig.1a. The crystal used in these experiments is roughly 1x 0.8x 2.7 mm in dimension. The crystal of this size was chosen due to the desire to measure low frequency thermal oscillations, which have been shown to arise only in large single crystals. \cite{Stern16} Single crystals of SmB$_6$ were grown using an aluminum flux method described in previous work.\cite{Kim12} 

Thermometry was performed via four probe resistance measurements of a RuO$_2$ thermometer mounted to a single surface of the studied crystal via thermally conducting varnish separated by a thin electrically insulating layer.The Vishay \RuO thermometer, as depicted in Fig.1b is primarily constructed from an alumina substrate with electrodes capping two ends that allow for electrical contact with the \RuO film which is protected by a thin epoxy layer. Due to the thick substrate sitting below the film resistor, the thermometer is mounted to the crystal such that the epoxy layer is in closest contact. Though the entire thermometer has a size on the order of that of the crystal, the \RuO film is only 0.8x1.15x0.020mm. RuO$_2$ was chosen as it has a high sensitivity in the vicinity of the transition of SmB$_6$ between bulk dominated and surface dominated conduction, a crucial transition for the generation of an oscillatory output voltage. In this regard we expect the thermometer to be quite sensitive to small changes in sample temperature. Four Manganin wires soldered to the thermometer were used as electrical contacts by which a small 1$\mu$A current was applied and a voltage measured. Manganin wires were chosen as they are poor thermal conductors at low temperatures ensuring the thermometer maximally thermally isolated from the thermal bath.

\begin{figure}
\label{Fig1}
\includegraphics[width=9cm,, height=10cm]{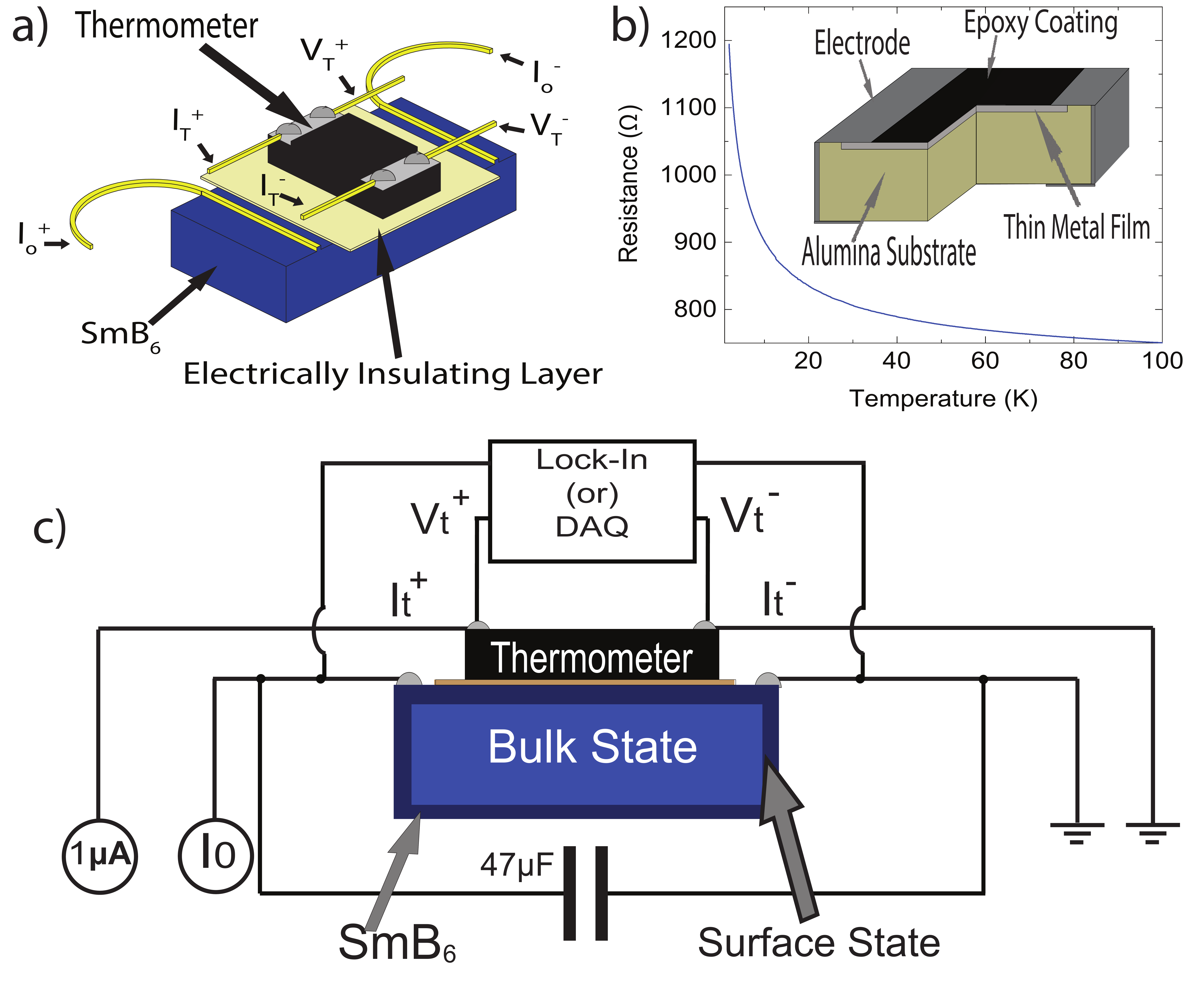}
\caption{(Color online) {\bf a,} A RuO$_2$ micro-thermometer is affixed via varnish to a thin electrically insulating layer on top of SmB$_6$. {\bf b,} Thermometer temperature dependent resistance {\bf insert} construction of RuO$_2$ thermometer {\b c,} Device measurement schematic in which either a Lock-In amplifier mesaures the thermometer voltage or a DAQ 9824.}
\end{figure}

Thermal measurements were performed via four probe method as described previously. The change in thermometer resistance was used in comparison with the temperature dependent resistance data taken upon cooling the thermometer to 2K as seen in Fig.1b in order to infer a thermometer temperature. Due to the requirement for temporally sensitive measurements of the thermometer temperature, it was of interest to first test the response time and magnitude of the thermometer junction. It is believed that a small thermal lag could be presented due to the heat transfer through the interfacial varnish and insulating layer. 

Testing this response was performed by measuring the voltages across the crystal and thermometer during which a small excitation current is switched on quickly to the crystal causing Joule heating. From this data both the crystal temperature and thermometer temperature were calculated using their respective temperature dependent resistance curves. This data as shown in Fig.2a shows a delay time for the thermometer below a measureable time window ($<$0.2ms). This short delay time suggests that no substantial phase shift should be introduced by the experimental design, thus allowing for reliable time dependent measurements within the frequency range studied. However, the time for equilibration of the thermometer temperature does lag behind that of the crystal. In fact half of the over-all change in temperature occurs in roughly 4ms as shown in Figure 2b, an enlargement of the boxed region in Fig 2a.
The 5ms scale appears to be in part due to the slow equilibration of the crystal shortly after the current is quickly turned on as shown by the gradual rise in temperature from 3.2K to 3.4K over the scale of roughly 10ms.  As such, we do not expect the thermometry to contribute substantially to any phase shift between the voltage oscillations and thermal oscillations, which should have a roughly constant relative phase of 90\textsuperscript{o}. Furthermore, at temperature saturation there is a difference in the temperatures between the crystal and thermometer, which can be seen in black in Fig. 2a. The saturation difference of roughly 25$\%$ was used to scale all measured thermometer temperatures to scale up to an estimated crystal temperature. It is thought that only a portion of the heat generated by Joule heating of SmB$_6$ successfully transfers to the thermometer, as some is lost to the insulating layer as well as through the Pt leads to the thermal bath contact to the crystal.”

\begin{figure}
\label{Fig2}
\includegraphics[width=9cm,, height=12cm]{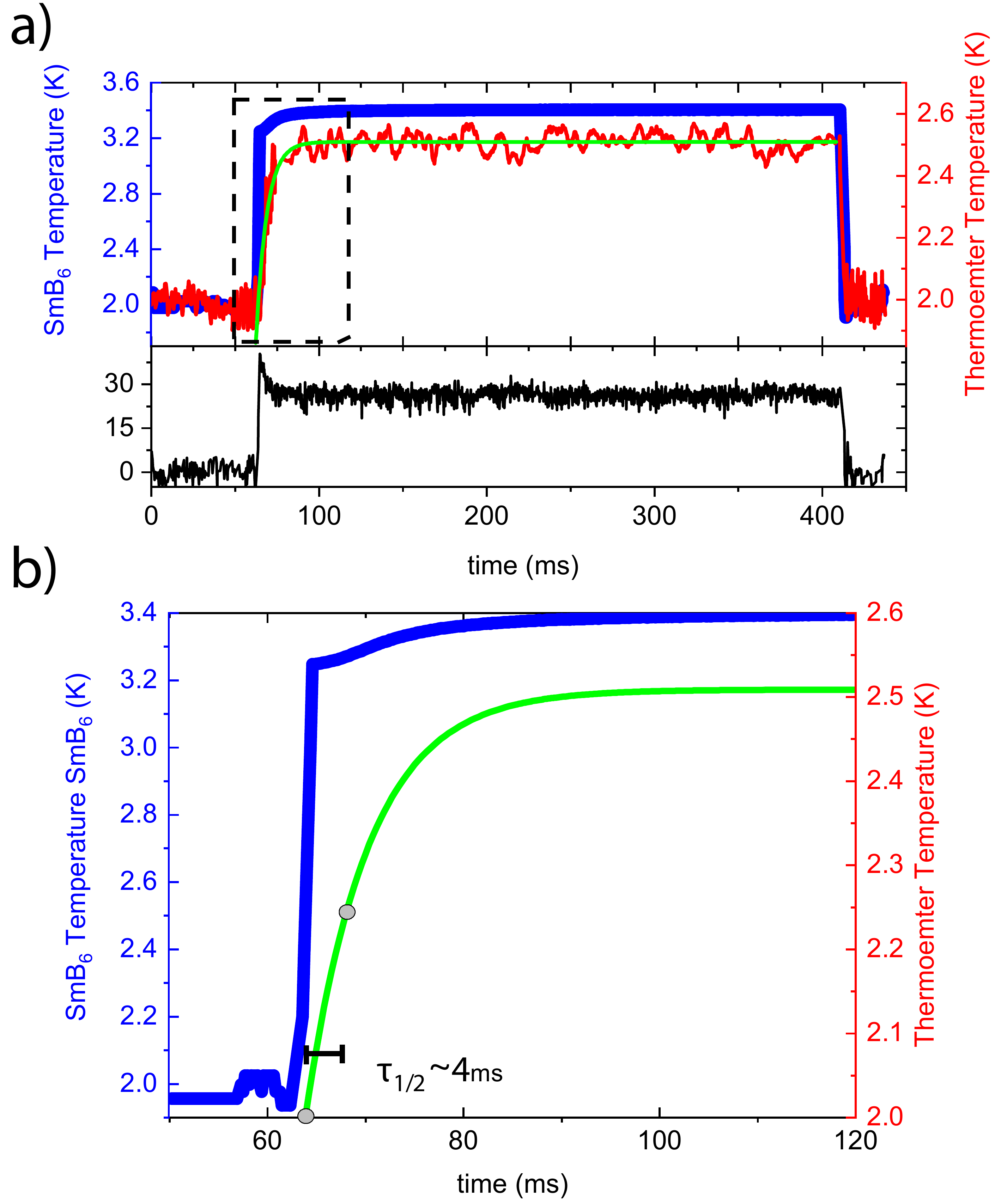}
\caption{(Color online) {\bf a,} Reponse of SmB$_6$ crystal (blue) and RuO$_2$ thermometer (red) to a square wave of current. The thermometer response was fit to an exponential function plotted in green. Shown in black is the percent difference bwteen the crystal tempearture and thermoemter temperature showing a peak when the current is applied that settles to roughly 25$\%$. {\bf b,} Englargement of the boxed region in Fig.2a showing the fitting curve and crystal response which reaches one half of its equilibrium temperature in 5ms. }
\end{figure}

Using the measurement scheme depicted in Fig.1c using the Lock-IN Amplifier, the measurement of the thermal oscillation frequency and amplitude can be carried out reliably, due in large to the rapid temperature response from the thermometer as previously discussed.  In this measurement design, the oscillatory voltage generated from the crystal under small DC bias is used as a reference signal for the measurement of the thermal oscillations. The data collected for the frequency and amplitude of these oscillations are shown in Fig. 3a,b.  We note that the surface temperature oscillation is less than 1.5 K, which agrees with our expectation. These data also behave consistently with simulations from previous works \cite{Stern16}. This method however lacks temporal resolution to observe thermal oscillation waveforms directly, requiring rapid data collection.

\begin{figure}
\label{Fig3}
\includegraphics[width=9cm,,height = 4cm]{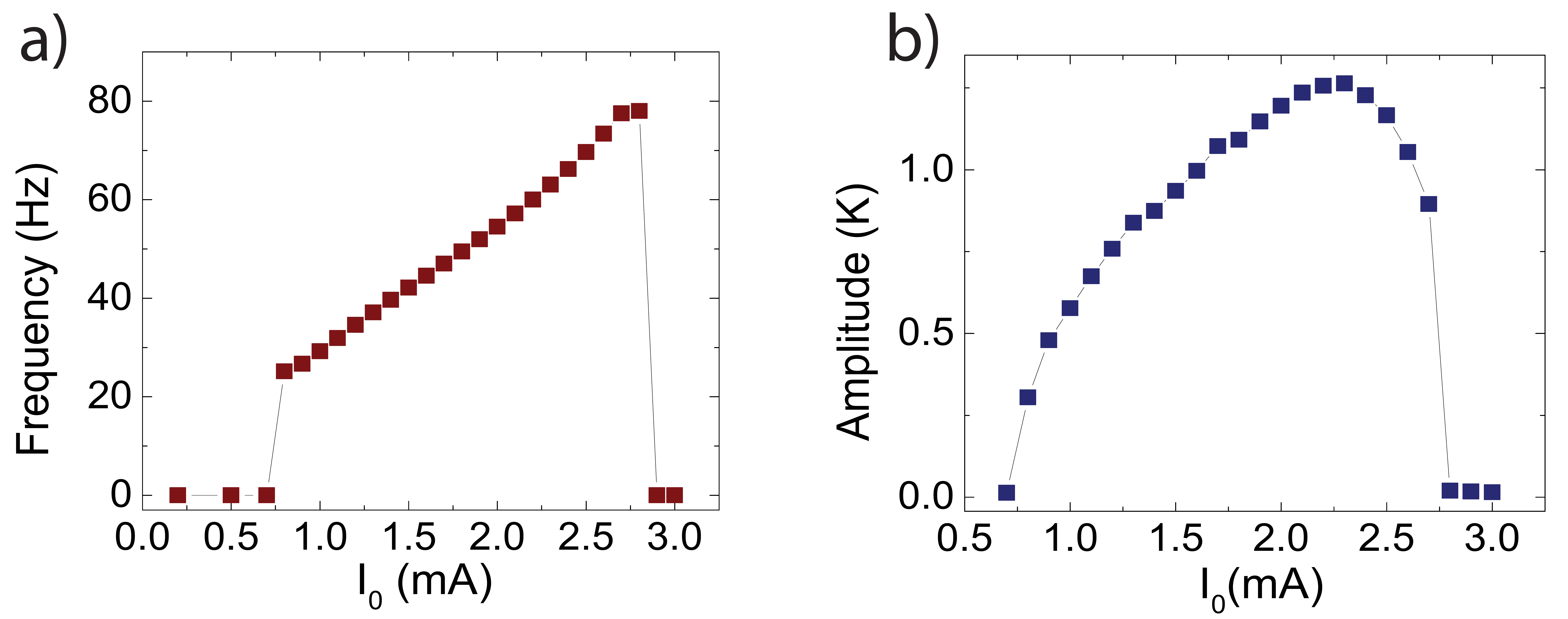}
\caption{(Color online) {\bf a,} As measured by a Lock-In amplifier, the frequency of thermal oscillations of the SmB$_6$ crystal as a function of applied DC bias. {\bf b,} Amplitude of thermal oscillations of SmB$_6$ crystals as a function of applied DC bias. The decrease sudden decrease in frequency and amplitude agree with previous simulations}
\end{figure}

A direct recording of the exact waveforms of thermal oscillation requires fast acquisition instruments such as a Digital Acquisition Device (DAQ). However, due to the lack of signal to noise amplification that is present in a Lock-In amplifier, thousands of periods of data must be collected to enable manual signal processing to observe signals with a low signal to noise ratio. A 9824 DAQ card was used to simultaneously measure the crystal oscillation voltage and thermometer voltage. For clarity we label all voltages and currents applied or measured from the thermometer as V$_t$ and I$_t$ while the crystal voltage and applied current are written as V$_o$ and I$_o$, respectively. All measured quantities are labeled in Fig. 1a on a cartoon of the previously described device. We constructed the measurement scheme as depicted in Fig. 1c in which all measurements were performed using the DAQ, which had a maximum measuring frequency of 4800Hz, which was set constant through all measurements. The DAQ was well suited for these measurements due to its voltage resolution of roughly 1$\mu$V. This method has the benefit of measuring both crystal voltage and thermometer voltage simultaneously, which allowed for the signal processing described in detail in this letter. The DAQ was used to measure the voltages across both the thermometer and the sample, which was placed in parallel to a 47 $\mu$F capacitor which was held at room temperature outside the cryostat. A Keithley 6200 Precision Current Source applied a 1 $\mu$A of DC current to the thermometer to prevent self-heating. This requisite small probing current resulted in a fairly low signal to noise ratio, making identification of the oscillatory voltage challenging. In using the DAQ to simultaneously measure the crystal voltage and the thermometer voltage, V$_t$, we were able to use V$_o$ as a trigger for the measurement of V$_t$. In this way the signal processing, i.e. averaging over thousands of periods of the data, was able to be performed without knowing the positions of the periods of the thermal oscillation. As predicted, the relative phase between thermal and voltage oscillations should constantly be 90\textsuperscript{o}.

\begin{figure}
\label{Fig4}
\includegraphics[width=9cm]{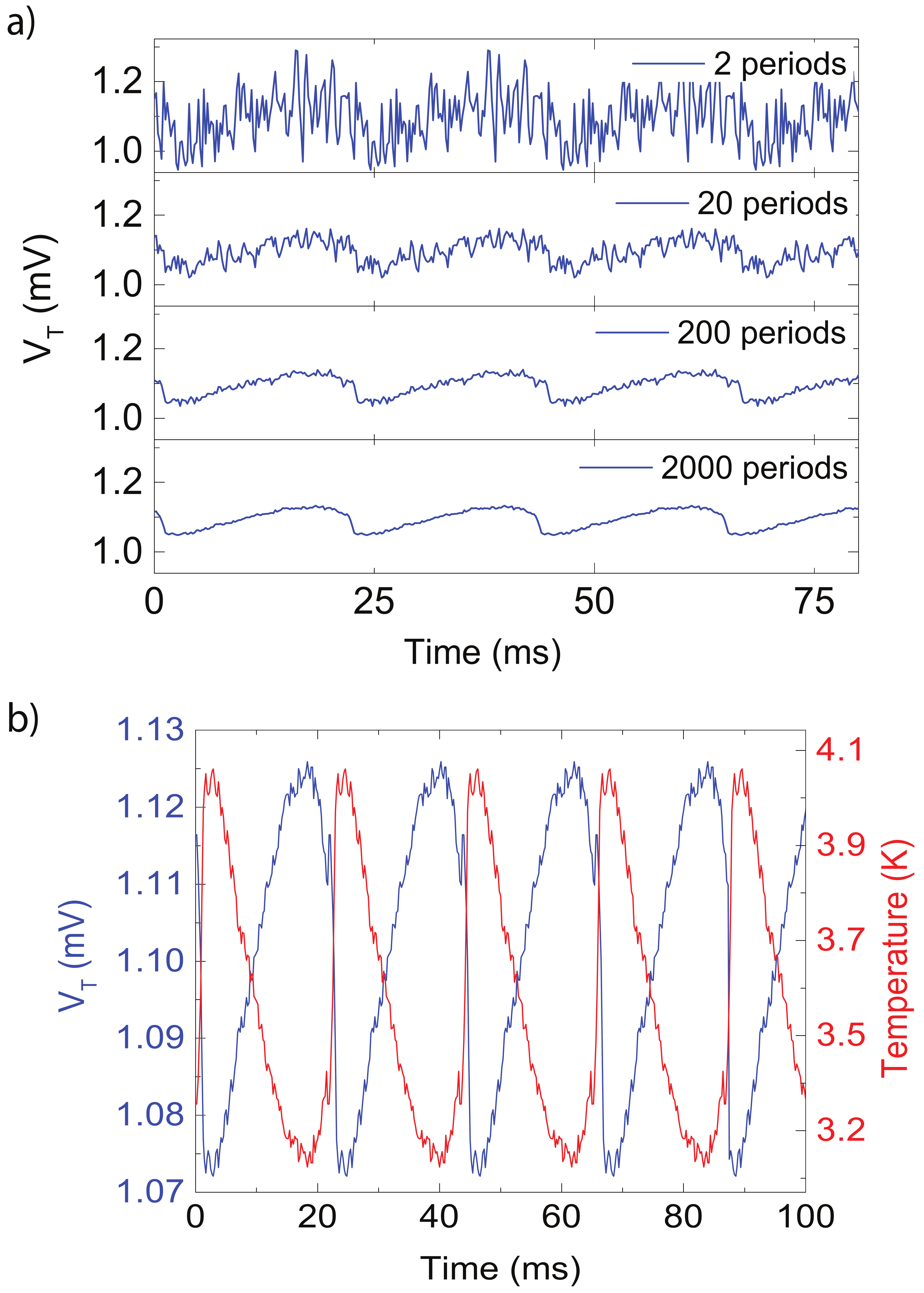}
\caption{(Color online) Waveform of surface state temperature oscillations revealed by DAQ measurements. {\bf a,} DAQ data after increased signal processing. The number of periods over which the data was averaged increases downward from 2 to 2000. {\bf b,} Raw V$_t$ data in blue and refined temperature oscillation in red all for sample with a 1.9mA bias. Processed V$_t$ data and RuO$_2$ resistance data were used to create temperature oscillation data. } 

\end{figure}

Signal processing involves three steps. First a custom script locates minima and maxima in the crystal voltage data V$_o$. Using these times, the data is separated into individual periods over which averaging will be performed. This averaging was typically performed over upwards of 2000 periods, though the exact number varied depending on the length of time over which data was collected. All crystal voltage data presented here is raw data, though most thermometer voltage data and temperature data has undergone this processing. The results of such processing can be seen in Fig. 4a, in which the periods averaged over increases from 2 periods to 2000 periods moving down the figure. The final result of the processing and temperature conversion can be seen in Fig. 4b. After completing the processing the data may be used to compare to the model as described previously. The processed thermal oscillation data and voltage oscillations data can be found in Fig.5. Note these have been plotted in comparison to curves generated by solving the proposed model in equation 1 which are in good agreement with the data taken via DAQ with I$_0$ = 1.9mA. Due to the simplicity of the proposed model, it is not surprising to observe some deviation in the waveform shapes between the model and experimental data.

\begin{figure}
\label{Fig5}
\includegraphics[width=9 cm, height=8cm]{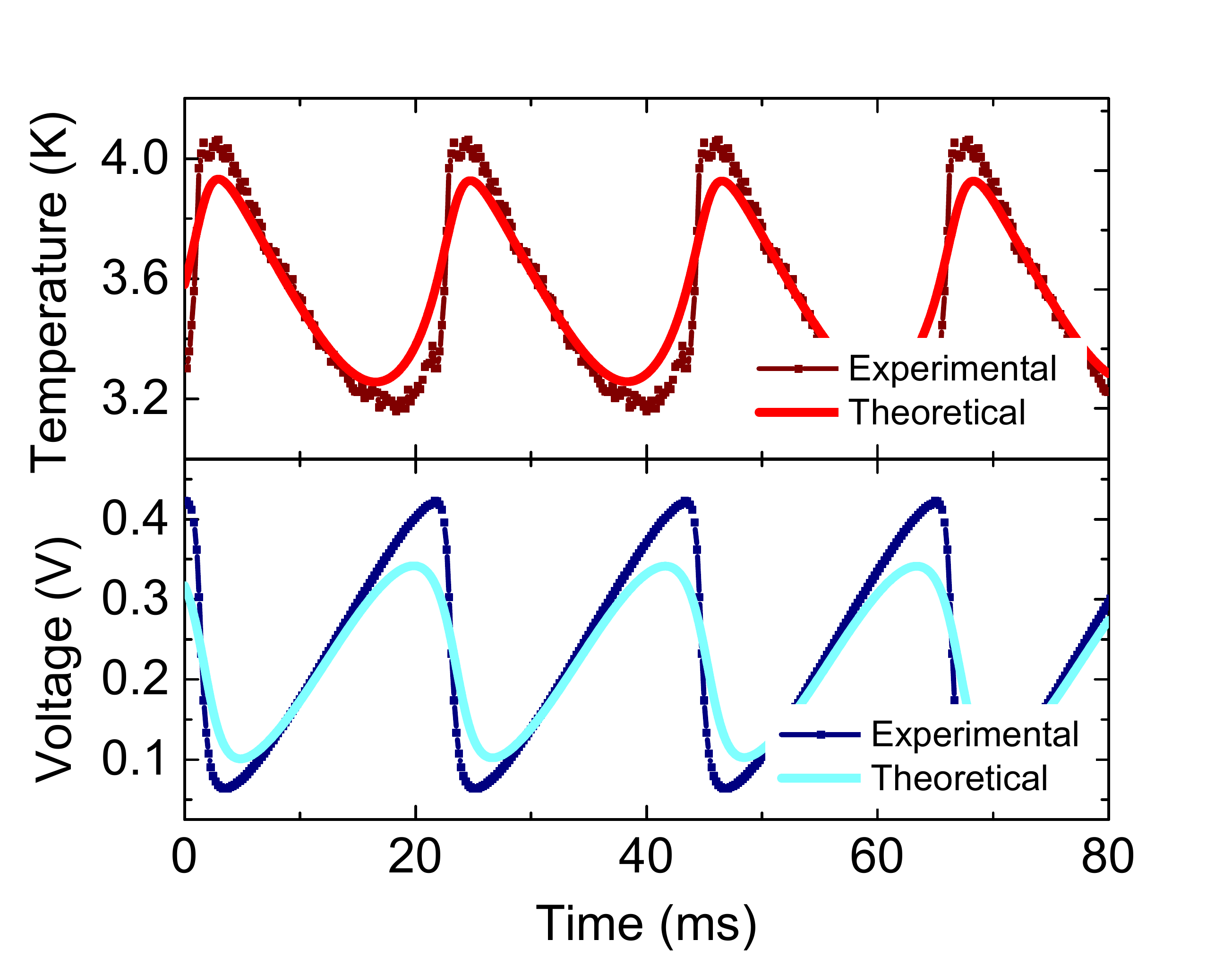}
\caption{(Color online) Comparing experimentally obtained temperature waveform with theoretical prediction. Temperature and crystal voltage oscillations plotted as red and blue connected points. Note that as predicted these waves appear to be roughly 90 degrees out of phase. Solid curves in red and blue represent data produced by the aforementioned nonlinear model using specific sample parameters.}

\end{figure}

The theoretical model and its regime diagram have been discussed in details in our previous paper~\cite{Stern16}. Here we briefly outline fitting of the current and temperature time dependencies, presented in Fig. 3. The model has four dimensionless controlling parameters and its fitting is not straightforward. Two of them $\delta=\Delta/T_B\approx19$ and $\rho=R_B^0/R_S\approx35$ can calculated directly since $R_S$, $R_B^0$ and $T_B$ have been measured in experiment. The other two $\Omega=C_H^0/\gamma C R_\mathrm{S}$ and $i_0=I_0 \sqrt{R_\mathrm{S}/\gamma T_B}$ depend on the heat transfer rate $\gamma$ and heat capacitance $C_\mathrm{H}^0$, which are difficult to estimate and measure. We will use them, as well as capacitance $C$ as fitting parameters. $\Omega$ determines the range of dimensionless external currents $i_0^\mathrm{min}$ and $i_0^\mathrm{max}$, where oscillations do appear. Their ratio $i_0^\mathrm{max}/i_0^\mathrm{min}=I_0^\mathrm{max}/I_0^\mathrm{min}$ defines it as $\Omega\approx0.46$. The last parameter $i_0$ can be calculated as is $i_0=I_0 i_0^\mathrm{max}/I_0^\mathrm{max}=I_0 i_0^\mathrm{min}/I_0^\mathrm{min}\approx0.95$. For this set of parameters we calculate the dimensionless period $T_0$ and fix the capacitance $C$, so that $T_\mathrm{exp}=C R_\mathrm{S} T_0 $. It results in $C_\mathrm{fit}=29\;\mu\hbox{F}$ while $C_\mathrm{exp}=47 \;\mu \hbox{F}$ which appear to be in modest agreement.  Now all parameters in equations (1) are determined, and their numerical integration results in theoretical curves in Fig.5 and Fig.6.

Deviations from the proposed model may stem from various origins. One such origin may be the previously reported non-trivial magnetic behavior manifesting as either ferromagnetic domains confined to the surface\cite{Nakajima1} , or as fluctuating magnetic fields present in the bulk\cite{Biswas1}. It is believed that the temperature range of interest in this study precludes the existence of these magnetic domains, thus no contribution to the model is expected. However the aforementioned magnetic fluctuations may play a role in this transport data if it contributes in a way to the heat capacity of SmB$_6$ or the thermal linkage to the bath. However, due to the small magnitude of those fluctuations as well as the sub-$\mu$s time scale it is unclear what affect they may have on these time dependent thermal properties. Generally, the proposed model in equation 1, relies very generally on conservation of energy and charge. Specific scattering terms that occur in transport are accounted for entirely by the surface or bulk resistance and specific heat. Any interaction that may lead to modifications of those conservation laws could in theory modify the model to suit more complex system behavior.

\begin{figure}
\label{Fig6}
\includegraphics[width=9cm]{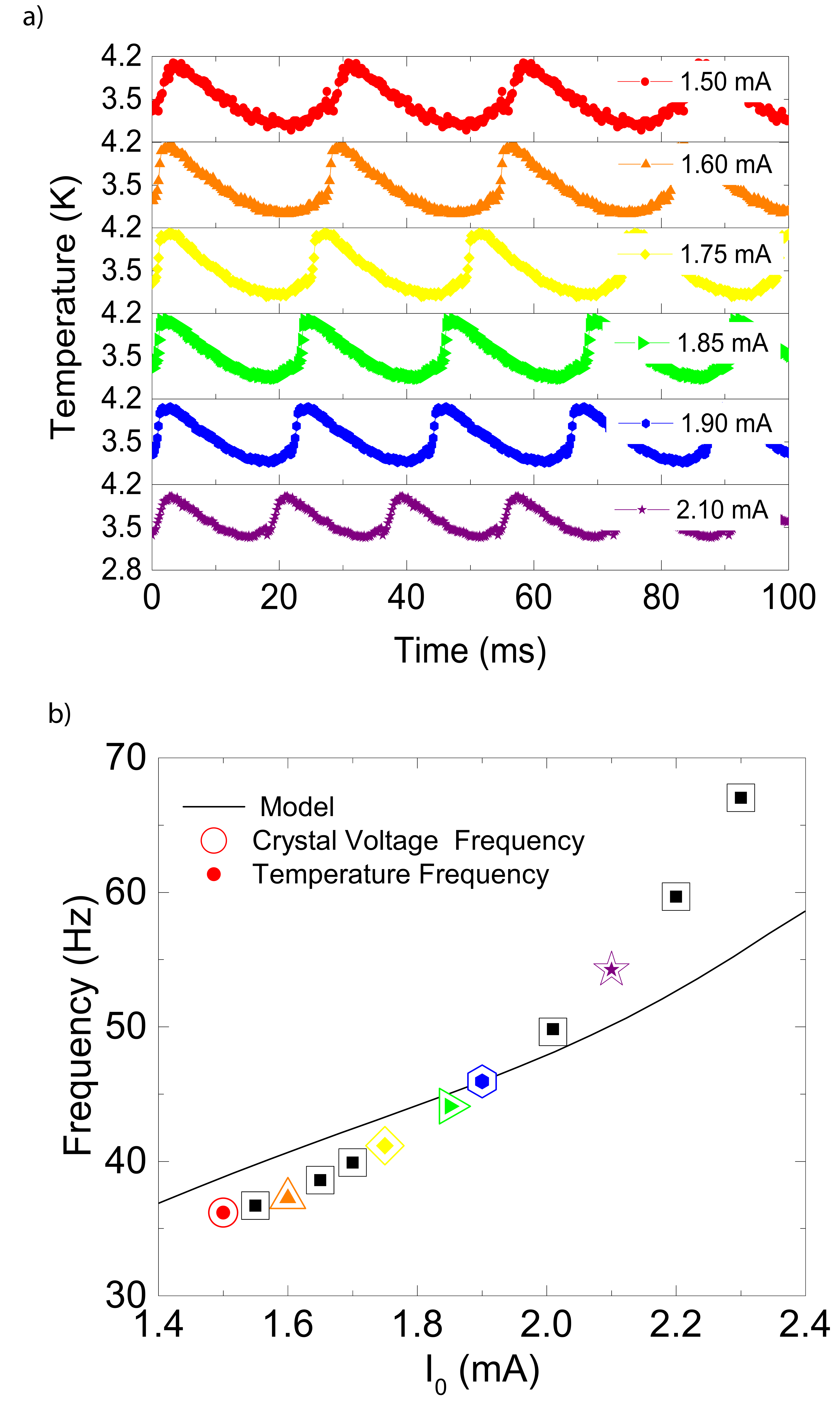}
\caption{(Color online) DC drive current dependence. {\bf a,} Waveforms of selected applied currents I$_o$ ranging from 1.5mA to 2.1mA. {\bf b,} Center frequency of crystal voltage oscillations as calculated from FFT using raw data. Colored symbols correspond to curves in Fig. 4a. {\bf c,} Center frequency of thermal oscillations as calculated from FFT of raw thermometer voltage data. Colored symbols correspond to curves in Fig. 4a. }
\end{figure} 

Select waveforms for six currents can be seen in Fig. 6a under various values of I$_o$. The frequencies of crystal voltage and thermal oscillations can be seen in Fig. 6b. In order to measure the frequency of the crystal voltage and temperature oscillations, FFT spectra calculated from raw data were fit using a Lorentzian to identify the center frequency and peak width. The full width at half max was used to estimate an upper bound in the uncertainty in the frequency, which in all cases were negligibly small. For the sake of comparison, the current dependent frequency plot can be constructed from the data collected by the DAQ and processed with the aforementioned method. Note that using the model described above, using parameters from this particular sample as discussed previously we were able to construct the dashed curve of frequencies as a function of applied current as seen in Fig. 6b in black. It should be noted that the theoretical data was calculated to match the frequency of the 1.9mA point. Interestingly, it appears that the experimental data is more sensitive to the crystal bias than the model had suggested, though both are in good agreement.

In summary, we have experimentally demonstrated the surface state thermal oscillation in SmB$_6$ oscillator devices. The observed amplitude, phase and waveform all agree well with the proposed theoretical model \cite{Stern16}, and will allow the development of THz SmB$_6$ oscillators with quantitative theoretical guidance. 

Acknowledgements: This work is supported by NSF grants DMR-1350122 and DMR-1708199

\end{document}